\documentclass[aps,prd,amsmath,amssymb,preprintnumbers,preprint,nofootinbib,a4paper,11pt]{revtex4-1}
\pdfoutput=1
\usepackage{graphicx}
\graphicspath{{./figures/}}
\usepackage{url}
\usepackage[bookmarks, pagebackref=false]{hyperref}
\usepackage[usenames,dvipsnames]{xcolor}
\definecolor{orange}{cmyk}{0,0.5,1,0}
\definecolor{rossoCP3}{cmyk}{0,.88,.77,.40}
\definecolor{graa}{rgb}{0.8,0.8,0.8}
\definecolor{blaa}{rgb}{0.2,0.2,0.6}
		\hypersetup{
			colorlinks, 
			bookmarksopen, 
			bookmarksnumbered,
			citecolor=blaa, 		
			linkcolor=rossoCP3,	
			urlcolor=rossoCP3,			
			}
\usepackage{amsthm}
\usepackage{bm}
\usepackage{bbm}
\usepackage{pxfonts}

\usepackage{placeins}
\usepackage{xspace}
\usepackage{cancel} 

\usepackage{slashed}
\usepackage[caption=false, labelformat=simple, listofformat=subsimple, labelfont=default, margin=5pt, justification=raggedright]{subfig}

	\makeatletter
		\renewcommand{\p@subfigure}{}
	\makeatother

\usepackage{natbib}

\newcommand{\beq}{\begin{eqnarray}}
\newcommand{\eeq}{\end{eqnarray}}

\newcommand{\bmp}{\noindent\begin{minipage}{16cm}}
\newcommand{\emp}{\end{minipage}\vskip 7mm} 

\newcommand{\eff}{\textrm{eff}}

\newcommand{\betafunction}{$\beta$~function\xspace}
\newcommand{\betafunctions}{$\beta$~functions\xspace}

\newcommand{\MSbar}{$\overline{\textrm{MS}}$}
\renewcommand{\d}{\textrm{d}}

\def\lsim{\mathrel{\rlap{\lower4pt\hbox{\hskip1pt$\sim$}}
    \raise1pt\hbox{$<$}}}                
\def\gsim{\mathrel{\rlap{\lower4pt\hbox{\hskip1pt$\sim$}}
    \raise1pt\hbox{$>$}}}                

\begin{document}
\title{\texorpdfstring{\Large\color{rossoCP3}  Standard Model Vacuum Stability and Weyl Consistency Conditions}{}}
\author{Oleg Antipin}
\email{antipin@cp3.dias.sdu.dk} 
\author{Marc Gillioz}
\email{gillioz@cp3.dias.sdu.dk} 
\author{Jens Krog}
\email{krog@cp3.dias.sdu.dk}  
\author{Esben M\o lgaard}
\email{molgaard@cp3.dias.sdu.dk}  
\author{Francesco Sannino}
\email{sannino@cp3.dias.sdu.dk}  
 \affiliation{
{ \color{rossoCP3}  \rm CP}$^{\color{rossoCP3} \bf 3}${\color{rossoCP3}\rm-Origins} \& the Danish Institute for Advanced Study {\color{rossoCP3} \rm DIAS},\\ 
University of Southern Denmark, Campusvej 55, DK-5230 Odense M, Denmark.
}

\begin{abstract}

At high energy the standard model possesses conformal symmetry at the classical level. This is reflected at the quantum level by relations between the different \betafunctions of the model. These relations are known as the Weyl consistency conditions. We show that it is possible to satisfy them order by order in perturbation theory, provided that a suitable coupling constant counting scheme is used.
As a direct phenomenological application, we study the stability of the standard model vacuum at high energies and compare with previous computations violating the Weyl consistency conditions. 

\vspace{0.5cm}

\noindent
{ \footnotesize  \it Preprint: CP$^3$-Origins-2013-020 \& DIAS-2013-20}

\end{abstract}

\maketitle

\section{Introduction}

With the discovery of the Higgs boson and the measurement of its mass, all the parameters of the standard model are now determined by experimental data with reasonable accuracy. Since the standard model is a renormalizable theory, its validity can be extended to any energy scale. Although the theory, as it is, is certainly incomplete, most notably because  it does not account for the experimental evidence of neutrino masses and missing mass in the universe, the time has come to ask whether the standard model is a valid description of physical phenomena up to the scale where gravity becomes strongly coupled, i.e. the Planck scale. For this possibility to be realized, it is necessary for the theory to possess a stable (or long-lived metastable) vacuum through the entire energy range.

Intriguingly, the measured values of the standard model couplings at the electroweak scale seem to lead to a peculiar situation near the Planck scale: the standard model is very close to the boundary between a stable and an unstable vacuum. A precise determination of the fate of the standard model relies, of course, on a precise knowledge of its couplings. The most relevant of these, both in terms of impact and uncertainty, are the top-Yukawa coupling, related to its mass measurement, and the Higgs self-coupling measured indirectly via its mass \cite{EliasMiro:2011aa}. At the same time, however, relevant uncertainties come also from the theory side. For example, since the determination of the vacuum stability in the standard model requires the running of its couplings over 16 orders of magnitude, theorists should provide the most precise computations possible. Much effort has been put into this task, and recently the three-loop \betafunctions in the \MSbar\ scheme have been obtained for the gauge \cite{Mihaila:2012fm, Bednyakov:2012rb}, Yukawa \cite{Bednyakov:2012en} and Higgs quartic coupling \cite{Chetyrkin:2012rz, Chetyrkin:2013wya, Bednyakov:2013eba}. At the same time, the determination of the \MSbar\ parameters at the electroweak scale has been  upgraded to the next-to-leading order \cite{Bezrukov:2012sa, Degrassi:2012ry, Alekhin:2012py}, so that state-of-the-art computations are now possible using two-loop matching and three-loop running of the coupling constants, together with a computation of the Higgs effective potential at two-loop order and including resummation of logarithms \cite{Degrassi:2012ry, Masina:2012tz}. The results seem to indicate that the standard model lives in a tiny region of metastability.

In this paper we will argue that the running of the couplings should be determined, together with its implication for the vacuum stability of the theory, via a more consistent framework. While it is true that the three-loop \betafunctions of the gauge, Yukawa and quartic couplings describe the most accurate determination of the running for each coupling separately, we will show that the correlated running of the different couplings requires a different counting in loops. The starting point of our argument is the conformal symmetry of the standard model. In the energy regime $E \gg v \approx 246$~GeV, the only operator with a dimensionful coupling, $H^\dag H$, can be neglected and the theory possesses a conformal symmetry at the classical level. This symmetry is, of course, broken at the quantum level, even in the absence of an explicit mass term for the Higgs field. This leads to a renormalization group (RG) flow. Nevertheless, not all consequences of the conformal symmetry are lost in the quantum theory. Among the remnants of the conformal symmetry, there exist a set of relations between the \betafunctions of the different couplings, known as the Weyl consistency conditions \cite{Osborn:1989td, Jack:1990eb, Osborn:1991gm}. These relations are made explicit below, and relate the coefficients of the \betafunctions at different loop orders. The one-loop running of the Higgs quartic coupling is, for instance, tightly related to the two-loop running of the top Yukawa and the three-loop running of the gauge couplings. The state-of-the-art computations, going to the three-loop order in gauge, Yukawa and quartic couplings (which we shall denote by 3-3-3 counting), explicitly break the Weyl consistency conditions. Establishing a {\it precision running} of the standard model couplings certainly requires that the conformal symmetry of the model is respected.

A consistent counting of the different loops contributing to the various \betafunctions respecting the conformal symmetry is possible. As we shall show, one has to consider the gauge \betafunctions at one loop order above the Yukawa one, and two orders above the quartic one. With the current knowledge of the \betafunctions, this permits only a counting of the type 3-2-1, i.e. at three-loop in gauge, two in Yukawa and one in the quartic coupling. If one wants to consider the running of the Higgs self-interactions at the three-loop order, a 5-4-3 counting is required, and therefore the knowledge of the gauge and Yukawa \betafunctions to higher precision is needed.  

The content of this work is organised as follows. In Section~\ref{sec:weylconsistencyconditions}, the Weyl consistency conditions on the \betafunctions are reviewed and they are shown to hold in the standard model in Section~\ref{sec:SM}.  A perturbative counting consistent with the conformal symmetry is established in Section~\ref{sec:counting}.  The explicit analysis of the vacuum stability according to the consistent 3-2-1 counting scheme is then presented in Section~\ref{sec:321}, where we also compare our results to the existing ones. We offer our conclusions in Section~\ref{sec:conclusions}.

\section{Conformal symmetry and Weyl consistency conditions}
\label{sec:weylconsistencyconditions}

We first review briefly the derivation of the Weyl consistency conditions. For a complete overview in two and four dimensions we refer to the seminal work of Osborn \cite{Osborn:1991gm}.  Consider a conformal theory augmented by a set of dimension four marginal operators breaking the conformal symmetry at the quantum level. With each such operator $\mathcal{O}^i$ is associated a coupling $g_i$, so that the Lagrangian of the theory can be summarized as
\begin{equation}
	\mathcal{L} = \mathcal{L}_{CFT} + g_i \mathcal{O}^i \, ,
\end{equation}
where $\mathcal{L}_{CFT}$ contains the kinetic terms for the fields of the theory. If one  disregards the mass term operator of the Higgs field, the standard model belongs to this class of models. In this case, the set of couplings $\{ g_i \}$ consists of the hypercharge, weak and strong couplings, the top-Yukawa and the Higgs quartic interaction, $\{ g_1, g_2, g_3, y_t, \lambda\}$. The subleading Yukawa interactions can safely be neglected for the purpose of this work.

Keeping track of the classical conformal symmetry after the theory has been renormalized is not straightforward. A convenient way to do so is to promote, at first, the couplings to functions of space-time, i.e.~$g_i = g_i(x)$, and to work in an arbitrary curved background. Under these assumptions, a conformal transformation of the space-time metric $\gamma_{\mu\nu} \to e^{2 \sigma(x)} \gamma_{\mu\nu}$ is partially compensated by a change in the renormalized coupling as $g_i(\mu) \to g_i(e^{-\sigma(x)} \mu)$, up to a number of terms that vanish in the limit of flat space-time and constant couplings. This can be explicitly encoded in the infinitesimal variation of the generating functional $W = \log\left[\int \mathcal{D}\Phi \, e^{i \int \d^4x \mathcal{L}} \right]$, parametrized as
\begin{equation}
	\Delta_\sigma W \equiv \int \d^4x \, \sigma(x)
		\left( 2 \gamma_{\mu\nu} \frac{\delta W}{\delta \gamma_{\mu\nu}}
			- \beta_i \frac{\delta W}{\delta g_i} \right)
		= \sigma \left( a E(\gamma) 
			+ \chi^{ij} \partial_\mu g_i \partial_\nu g_j G^{\mu\nu} \right)
			+ \partial_\mu \sigma w^i \, \partial_\nu g_i G^{\mu\nu} 
			+ \ldots
	\label{eq:weylanomaly}
\end{equation}
where $a$, $\chi^{ij}$ and $\omega^i$ are functions of the renormalized couplings, $\beta_i$ denotes the \betafunction associated to the coupling $g_i$, $E(\gamma) = R^{\mu\nu\rho\sigma} R_{\mu\nu\rho\sigma} - 4 R^{\mu\nu}R_{\mu\nu} + R^2$ is the Euler density and $G^{\mu\nu} = R^{\mu\nu} - \frac{1}{2	} \gamma^{\mu\nu} R$ is the Einstein tensor. The right-hand side of the equation contains all possible dimension-four Lorentz scalars constructed out of the metric and derivatives of the couplings, $\partial_\mu g_i$, and only the three terms relevant to our discussion have been shown here. The functions $a$, $\chi^{ij}$ and $\omega^i$ are completely determined by the theory and can be explicitly computed in a perturbative expansion in the couplings $g_i$. The essence of the Weyl consistency conditions is that these functions are not independent of each other. In particular, the Weyl anomaly expressed by Eq.~\eqref{eq:weylanomaly} is of abelian nature, and therefore must satisfy
\begin{equation}
	\Delta_\sigma \Delta_\tau W = \Delta_\tau \Delta_\sigma W.
	\label{eq:abeliananomaly}
\end{equation}
This equation gives a number of relations between the terms to the right-hand side of Eq.~\eqref{eq:weylanomaly}, among which\footnote{In the presence of dimension-three currents in the theory, a few subtleties arise in the derivation of the consistency condition \eqref{eq:consistencycondition}, and the $\beta_i$ must be replaced by a different quantity denoted by $B_i$ in Refs.~\cite{Fortin:2012hn, Luty:2012ww}. $\beta_i$ and $B_i$ agree however at the lowest orders in perturbation theory and we will not make a distinction between them in this work.}
\begin{equation}
	\frac{\partial \tilde{a}}{\partial g_i} = \left(-\chi^{ij} + \frac{\partial w^i}{\partial g_j} - \frac{\partial w^j}{\partial g_i} \right) \beta_j \, ,
	\label{eq:consistencycondition}
\end{equation}
where we have defined $\tilde{a} \equiv a - w^i \beta_i$. From this equation it follows that $\frac{d}{d\mu} \tilde{a} = - \chi^{ij} \beta_i \beta_j$, so that $\tilde{a}$ is monotonically decreasing along the RG flow, provided that $\chi$ is a positive definite matrix. $\chi$~is indeed positive definite at lowest order in perturbation theory~\cite{Jack:1990eb}, however not necessarily beyond this order (see e.g.~Ref.~\cite{Antipin:2013pya}). Establishing the positivity of $\chi$ beyond perturbation theory would immediately prove the $a$-theorem~\cite{Cardy:1988cwa} and the irreversibility of the RG flow in four dimensions.\footnote{Using analyticity arguments, it was shown recently that the function $\tilde{a}$ in the UV is bigger than in the IR \cite{Komargodski:2011vj, Komargodski:2011xv}. However, this method does not permit to draw any conclusions on the behaviour of $\tilde{a}$ along the RG flow.} We stress that equation \eqref{eq:consistencycondition} relies neither on the space dependence of the couplings nor the space-time metric. Henceforth we will work in ordinary Minkowski background. 

For a generic gauge-Yukawa theory, the function $w^i$ turns out to be an exact one-form at the lowest orders in perturbation theory \cite{Jack:1990eb}, so that the terms involving derivatives of $w^i$ cancel out, and we will use in the following the simplified consistency condition
\begin{equation}
	\frac{\partial \tilde{a}}{\partial g_i} = -\beta^i \, ,
	\hspace{2cm}
	\beta^i \equiv \chi^{ij} \beta_j.
	\label{eq:truncatedconsistencycondition}
\end{equation}
$\chi^{ij}$ can be seen as a metric in the space of couplings, used in this case to raise and lower latin-indices. The fact that all \betafunctions can be derived from the same quantity $\tilde{a}$ has profound implications. The flow generated by the modified \betafunctions $\beta^i$ is a gradient flow, implying in particular
\begin{equation}
	\frac{\partial \beta^j}{\partial g_i} = \frac{\partial \beta^i}{\partial g_j} \, ,
	\label{eq:integrabilitycondition}
\end{equation}
which gives relations between the \betafunctions of different couplings.
These consistency conditions can be used as a check of a known computation, but could, in principle, also be used to determine some unknown coefficients at a higher loop order in perturbation theory.

\section{The Weyl consistency conditions in the standard model}
\label{sec:SM}

We now specialize the above conditions to the important case of the standard model of particle interactions. The couplings we consider are the gauge couplings, the top-Yukawa and the quartic interaction of the Higgs field. Due to the nature of the perturbative corrections it is convenient to redefine the coupling set $\{  g_i \}$ as  $\{ \alpha_1, \alpha_2, \alpha_3, \alpha_t, \alpha_\lambda \}$, where
\begin{equation}
	\alpha_{1} = \frac{g_1^2}{(4 \pi)^2},
	\quad
	\alpha_{2} = \frac{g_2^2}{(4 \pi)^2},
	\quad
	\alpha_{3} = \frac{g_3^2}{(4 \pi)^2},
	\quad
	\alpha_t = \frac{y_t^2}{(4 \pi)^2},
	\quad
	\alpha_\lambda = \frac{\lambda}{(4 \pi)^2} \ .
	\label{eq:alphas}
\end{equation}
As explained above, $g_1, g_2, g_3$ are the $U(1)_Y$, $SU(2)_{\rm{w}}$ and $SU(3)_c$ gauge couplings respectively. Similarly, we denote by $\beta_1, \beta_2, \beta_3, \beta_t$ and $\beta_\lambda$ their respective \betafunction, defined as $\beta_{i}\equiv \mu^2 \frac{d\alpha_i}{d\mu^2}$. At leading order in the couplings, the matrix $\chi$ is diagonal, and reads \cite{Jack:1990eb}
\begin{equation}
	\chi = \textrm{diag}\left(\frac{1}{ \alpha_1^2}\ , \frac{3}{\alpha_2^2} \ ,
		 \frac{8}{\alpha_3^2}\ , \frac{2}{\alpha_t}\ , 4
		\right) \ .
	\label{eq:metric}
\end{equation} 
One finds that any gauge  $\beta^g$ compared  to the original $\beta_g$ features two powers fewer in $\alpha_g$; the Yukawa $\beta^t$ is related to $\beta_t$ with one less power of $\alpha_t$ while $\beta^{\lambda}$  carries the same powers in $\alpha_\lambda$ as $\beta_\lambda$. 

The condition (\ref{eq:integrabilitycondition}) therefore plays an important role, since it relates coefficients of different \betafunctions at different loop order. Explicitly, the lowest order consistency conditions that we obtain are
\begin{eqnarray}
	2 \frac{\partial}{\partial\alpha_t} \beta_\lambda
		& = & \frac{\partial}{\partial\alpha_\lambda} \left( \frac{\beta_t}{\alpha_t} \right) + \mathcal{O}\left( \alpha_i^2 \right)
		\label{eq:consistencycondition:lambda:t} \\
	4 \frac{\partial}{\partial\alpha_1} \beta_\lambda
		& = & \frac{\partial}{\partial\alpha_\lambda} \left( \frac{\beta_1}{\alpha_1^2} \right) + \mathcal{O}\left( \alpha_i^2 \right)
		\label{eq:consistencycondition:lambda:1} \\
	\frac{4}{3} \frac{\partial}{\partial\alpha_2} \beta_\lambda
		& = & \frac{\partial}{\partial\alpha_\lambda} \left( \frac{\beta_2}{\alpha_2^2} \right) + \mathcal{O}\left( \alpha_i^2 \right)
		\label{eq:consistencycondition:lambda:2} \\
	2 \frac{\partial}{\partial\alpha_1} \left( \frac{\beta_t}{\alpha_t} \right)
		& = & \frac{\partial}{\partial\alpha_t} \left( \frac{\beta_1}{\alpha_1^2} \right) + \mathcal{O}\left( \alpha_i^2 \right)
		\label{eq:consistencycondition:t:1} \\
	\frac{2}{3} \frac{\partial}{\partial\alpha_2} \left( \frac{\beta_t}{\alpha_t} \right)
		& = & \frac{\partial}{\partial\alpha_t} \left( \frac{\beta_2}{\alpha_2^2} \right) + \mathcal{O}\left( \alpha_i^2 \right)
		\label{eq:consistencycondition:t:2} \\
	\frac{1}{4} \frac{\partial}{\partial\alpha_3} \left( \frac{\beta_t}{\alpha_t} \right)
		& = & \frac{\partial}{\partial\alpha_t} \left( \frac{\beta_3}{\alpha_3^2} \right) + \mathcal{O}\left( \alpha_i^2 \right)
		\label{eq:consistencycondition:t:3}\\
	\frac{1}{3} \frac{\partial}{\partial\alpha_2} \left( \frac{\beta_1}{\alpha_1^2} \right)
		& = & \frac{\partial}{\partial\alpha_1} \left( \frac{\beta_2}{\alpha_2^2} \right) + \mathcal{O}\left( \alpha_i^2 \right)
		\label{eq:consistencycondition:1:2} \\
	\frac{1}{8} \frac{\partial}{\partial\alpha_3} \left( \frac{\beta_1}{\alpha_1^2} \right)
		& = & \frac{\partial}{\partial\alpha_1} \left( \frac{\beta_3}{\alpha_3^2} \right) + \mathcal{O}\left( \alpha_i^2 \right)
		\label{eq:consistencycondition:1:3} \\
	\frac{3}{8} \frac{\partial}{\partial\alpha_3} \left( \frac{\beta_2}{\alpha_2^2} \right)
		& = & \frac{\partial}{\partial\alpha_2} \left( \frac{\beta_3}{\alpha_3^2} \right) + \mathcal{O}\left( \alpha_i^2 \right)
		\label{eq:consistencycondition:2:3}
\end{eqnarray}

We can now proceed to test these relations for the  standard model \betafunctions.  We take them from Ref.~\cite{Holthausen:2011aa, Chetyrkin:2012rz, Mihaila:2012fm}, without using the $SU(5)$ normalisation for the hypercharge:
\begin{widetext}
\begin{align}
  \beta_1 &= 
  2\alpha_1^2 \Bigg\{
    \frac{1}{12} + \frac{10n_G}{9} 
    + \left(\frac{1}{4} + \frac{95n_G}{54}\right) \alpha_1
    + \underbrace{\left(\textcolor{RoyalPurple}{\frac{3}{4} + \frac{n_G}{2}}\right) \alpha_2}_{Eq.~\eqref{eq:consistencycondition:1:2}}
    + \underbrace{\textcolor{Brown}{\frac{22n_G}{9}} \alpha_3}_{Eq.~\eqref{eq:consistencycondition:1:3}} 	+ \left(  \frac{163}{1152} -\frac{145 n_G}{81} - \frac{5225 n_G^2}{1458} \right) \alpha_1^2 
    \nonumber\\&\mbox{}
    + \left(  \frac{87}{64} - \frac{7 n_G}{72} \right)  \alpha_1\alpha_2
    - \frac{137n_G}{162} \alpha_1\alpha_3
    + \left( \frac{3401}{384} + \frac{83n_G}{36} - \frac{11n_G^2}{18}  \right)  \alpha_2^2
    + \left(\frac{1375n_G}{54} - \frac{242n_G^2}{81}\right) \alpha_3^2
    - \frac{n_G}{6} \alpha_2\alpha_3
    \nonumber\\&\mbox{}
    + \alpha_t \bigg[
      \underbrace{-\textcolor{cyan}{\frac{17}{12}}}_{Eq.~\eqref{eq:consistencycondition:t:1}}
      - \frac{2827}{576} \alpha_1
      - \frac{785}{64} \alpha_2
      - \frac{29}{6} \alpha_3
      + \left( \frac{113}{32} + \frac{101 n_t}{16}\right) \alpha_t
    \bigg]
    + \underbrace{\alpha_{\lambda} \bigg(
     \textcolor{green}{\frac{3}{4}}  \alpha_1 
      + \textcolor{green}{\frac{3}{4}} \alpha_2
      - \textcolor{green}{\frac{3}{2}} \alpha_{\lambda}
      \bigg)}_{Eq.~\eqref{eq:consistencycondition:lambda:1}}
    \Bigg\}\,,
  \nonumber\\
  \beta_2 &= 
  2\alpha_2^2 \Bigg\{
    -\frac{43}{12} + \frac{2n_G}{3} 
    + \underbrace{\left(\textcolor{RoyalPurple}{\frac{1}{4}} + \textcolor{RoyalPurple}{\frac{n_G}{6}}\right) \alpha_1}_{Eq.~\eqref{eq:consistencycondition:1:2}}
    + \left(-\frac{259}{12} + \frac{49n_G}{6}\right) \alpha_2
    + \underbrace{\textcolor{Gray}{2 n_G} \alpha_3}_{Eq.~\eqref{eq:consistencycondition:2:3}}
    + \left(  \frac{163}{1152} -\frac{35 n_G}{54} - \frac{55 n_G^2}{162} \right) \alpha_1^2
    \nonumber\\&\mbox{}
    + \left(  \frac{187}{64} + \frac{13 n_G}{24} \right) \alpha_1\alpha_2
    - \frac{n_G}{18} \alpha_1\alpha_3
    + \left(  -\frac{667111}{3456} + \frac{3206 n_G}{27} - \frac{415
      n_G^2}{54} \right)  \alpha_2^2
    \nonumber\\&\mbox{}
    + \frac{13 n_G}{2} \alpha_2\alpha_3
    + \left(\frac{125n_G}{6} - \frac{22n_G^2}{9}\right) \alpha_3^2
    \nonumber\\&\mbox{}
    + \alpha_t \bigg[
      \underbrace{-\textcolor{magenta}{\frac{3}{4}}}_{Eq.~\eqref{eq:consistencycondition:t:2}}
      - \frac{593}{192} \alpha_1
      - \frac{729}{64} \alpha_2
      - \frac{7}{2} \alpha_3
      + \left( \frac{57}{32} + \frac{45 n_t}{16}\right) \alpha_t
    \bigg]
    + \underbrace{\alpha_{\lambda} \bigg(
      \textcolor{blue}{\frac{1}{4}} \alpha_1 
      + \textcolor{blue}{\frac{3}{4}} \alpha_2
      - \textcolor{blue}{\frac{3}{2}} \alpha_{\lambda}
      \bigg)}_{Eq.~\eqref{eq:consistencycondition:lambda:2}}
    \Bigg\}\,,
  \nonumber\\
  \beta_3 &=
  2\alpha_3^2 \Bigg\{
    -\frac{11}{2} + \frac{2n_G}{3} 
    + \underbrace{\textcolor{Brown}{\frac{11n_G}{36}} \alpha_1}_{Eq.~\eqref{eq:consistencycondition:1:3}}
    + \underbrace{\textcolor{Gray}{\frac{3n_G}{4}} \alpha_2}_{Eq.~\eqref{eq:consistencycondition:2:3}}
    + \left(-51 + \frac{38n_G}{3}\right) \alpha_3
    + \left(  -\frac{65 n_G}{432} - \frac{605 n_G^2}{972}  \right) \alpha_1^2
    \nonumber\\&\mbox{}
    - \frac{n_G}{48} \alpha_1\alpha_2
    + \frac{77 n_G}{54} \alpha_1\alpha_3
    + \left(  \frac{241 n_G}{48} - \frac{11 n_G^2}{12}  \right)  \alpha_2^2 
    +\frac{7 n_G}{2} \alpha_2\alpha_3
    \nonumber\\&\mbox{}
    + \left(-\frac{2857}{4} + \frac{5033n_G}{18} - \frac{325n_G^2}{27}\right) \alpha_3^2
    + \alpha_t \bigg[
      \underbrace{ -\textcolor{orange}{1}}_{Eq.~\eqref{eq:consistencycondition:t:3}}
      - \frac{101}{48} \alpha_1
      - \frac{93}{16} \alpha_2
      - 20 \alpha_3
      + \left( \frac{9}{4} + \frac{21 n_t}{4}\right) \alpha_t
    \bigg]
    \Bigg\}\,,
    \nonumber\\
    \beta_{t}&= 2\alpha_t \Bigg\{ \frac{9}{4}\alpha_t
    		- \underbrace{\textcolor{orange}{4} \alpha_3}_{Eq.~\eqref{eq:consistencycondition:t:3}}
    		- \underbrace{\textcolor{cyan}{\frac{17}{24}} \alpha_1}_{Eq.~\eqref{eq:consistencycondition:t:1}}
    		- \underbrace{\textcolor{magenta}{\frac{9}{8}}\alpha_2}_{Eq.~\eqref{eq:consistencycondition:t:2}} 
    		+ \underbrace{\textcolor{red}{3} \alpha_{\lambda}^2 
    		- \textcolor{red}{6} \alpha_t\alpha_{\lambda}}_{Eq.~\eqref{eq:consistencycondition:lambda:t}}
    		-6\alpha_t^2+18\alpha_3\alpha_t
	    	\nonumber\\&\mbox{}
	    + \alpha_3^2\Big(-\frac{202}{3}+\frac{40n_G}{9}\Big)
	    + \alpha_t\Big(\frac{131}{32}\alpha_1+\frac{225}{32}\alpha_2\Big)
	    + \frac{1187}{432}\alpha_1^2-\frac{3}{8}\alpha_1\alpha_2
	    + \frac{19}{18}\alpha_1\alpha_3-\frac{23}{8}\alpha_2^2
	    + \frac{9}{2}\alpha_3\alpha_2\Bigg\}\,,
	    &\mbox{}\nonumber\\
    \beta_{\lambda}& = \underbrace{\textcolor{blue}{\frac{9}{16}} \alpha_2^2
    		- \textcolor{blue}{\frac{9}{2}} \alpha_\lambda \alpha_2}_{Eq.~\eqref{eq:consistencycondition:lambda:2}}
    		+\underbrace{\textcolor{green}{\frac{3}{16}}\alpha_1^2
    		- \textcolor{green}{\frac{3}{2}} \alpha_\lambda \alpha_1}_{Eq.~\eqref{eq:consistencycondition:lambda:1} } 
    		+\underbrace{\frac{\textcolor{green}{3}}{\textcolor{blue}{8}} \alpha_1 \alpha_2}_{Eqs.~(\ref{eq:consistencycondition:lambda:1}-\ref{eq:consistencycondition:lambda:2})}  
    		+ 12 \alpha_\lambda^2
    		+ \underbrace{\textcolor{red}{6} \alpha_\lambda \alpha_t
    		- \textcolor{red}{3} \alpha_t^2}_{Eq.~\eqref{eq:consistencycondition:lambda:t}} \ .
    	\label{eq:betafunctions}
\end{align}
\end{widetext}
Here $n_G$ is the number of generations which we set to $3$ and $n_t$ is the number of top quarks, i.e. one. 
Note that although we considered the gauge \betafunctions to three loops, we show only the two-loop top Yukawa and the one-loop Higgs quartic \betafunctions. This, as we will demonstrate momentarily, leads to a Weyl consistent expansion in the couplings up to $\mathcal{O}(\alpha_i^3)$. 

To help the reader immediately identify the terms in the \betafunctions  that must satisfy the Weyl consistency conditions given in Eqs.~(\ref{eq:consistencycondition:lambda:t}-\ref{eq:consistencycondition:2:3}), we have color-coded the contributions. Furthermore, beneath each relevant term we have noted the equation number of the Weyl consistency condition it refers to. Specifically, the red color is associated to Eq~\eqref{eq:consistencycondition:lambda:t}, green to Eq.~\eqref{eq:consistencycondition:lambda:1}, blue to Eq.~\eqref{eq:consistencycondition:lambda:2},  cyan to Eq.~\eqref{eq:consistencycondition:t:1}, magenta to Eq.~\eqref{eq:consistencycondition:t:2}, orange to Eq.~\eqref{eq:consistencycondition:t:3}, purple to Eq.~\eqref{eq:consistencycondition:1:2}, brown to Eq.~\eqref{eq:consistencycondition:1:3}, and finally gray to Eq.~\eqref{eq:consistencycondition:2:3}. Note that the term $\frac{3}{8}\alpha_1\alpha_2$ in $\beta_\lambda$ enters into both Eq.~\eqref{eq:consistencycondition:lambda:1} and Eq.~\eqref{eq:consistencycondition:lambda:2}.

This illustrates that the one-loop coefficients of the quartic $\beta_{\lambda}$-function is related to the two-loop coefficient of the Yukawa $\beta_t$-function, and to the three-loop \betafunctions of the electroweak gauge couplings. Restricting the computation to these orders, namely adopting a 3-2-1 loop counting in the gauge, Yukawa and quartic \betafunctions, corresponds to a truncation of the function $\tilde{a}$ at order $\alpha_i^3$. For illustration, we show the terms in the function $\tilde{a}$ which contribute to the one-loop quartic $\beta_{\lambda}$-function:
\begin{equation}
 -\tilde{a}= \ldots + \underbrace{\textcolor{blue}{\frac{9}{4}} \alpha_2^2 \alpha_\lambda
    		- \textcolor{blue}{9} \alpha_\lambda^2 \alpha_2}_{Eq.~\eqref{eq:consistencycondition:lambda:2}}
    		+\underbrace{\textcolor{green}{\frac{3}{4}}\alpha_1^2 \alpha_\lambda
    		- \textcolor{green}{3} \alpha_\lambda^2 \alpha_1}_{Eq.~\eqref{eq:consistencycondition:lambda:1} } 
    		+\underbrace{\frac{\textcolor{green}{3}}{\textcolor{blue}{2}} \alpha_1 \alpha_2\alpha_\lambda}_{Eqs.~(\ref{eq:consistencycondition:lambda:1}-\ref{eq:consistencycondition:lambda:2})}  
    		+ 16 \alpha_\lambda^3
    		+ \underbrace{\textcolor{red}{12} \alpha_\lambda^2 \alpha_t
    		- \textcolor{red}{12} \alpha_t^2\alpha_\lambda}_{Eq.~\eqref{eq:consistencycondition:lambda:t}} + \ldots 
\end{equation}

\section{A consistent perturbative expansion}
\label{sec:counting}

When considering terms in the \betafunctions of higher order than the ones present in Eq.~\eqref{eq:betafunctions}, one implicitly includes in the function $\tilde{a}$ terms of order $\alpha_i^4$ or higher. For instance, let us study a typical two-loop term in the quartic \betafunction ,
\begin{equation}
	\beta_\lambda = \ldots + \frac{45}{4} \alpha_2 \alpha_t \alpha_\lambda + \ldots \, .
	\label{eq:higherorderterm}
\end{equation}
It originates from a term of the form $\alpha_2 \alpha_t \alpha_\lambda^2$ in $\tilde{a}$, whose presence demands a term of order $\alpha_2 \alpha_t \alpha_\lambda^2$ in $\beta_t$, which only appears at the three-loop level, and another of order $\alpha_2^2 \alpha_t \alpha_\lambda^2$ in $\beta_2$, which is a four-loop term.\footnote{It is important to note, however, that one cannot simply infer the form of these terms directly from Eq.~\eqref{eq:higherorderterm}, since the metric $\chi^{ij}$ contains corrections of higher order in $\alpha_i$, not shown in Eq.~\eqref{eq:metric}. Some of these corrections have been computed in Ref.~\cite{Jack:1990eb}. } When truncating all \betafunctions to three loops, the absence of these terms explicitly violates the Weyl consistency conditions.

Our point in this paper is that for any analysis requiring the running of different couplings, a consistent perturbative expansion must be adopted in the function $\tilde{a}$, from which the counting of couplings in the various \betafunctions should then be derived. Truncating $\tilde{a}$ to order $\alpha_i^3$ corresponds to the 3-2-1 counting mentioned above. Similarly, truncations at order $\alpha_i^4$ or $\alpha_i^5$ in $\tilde{a}$ yield respectively the 4-3-2 or 5-4-3 Weyl-consistent countings. If, for instance, the NNLO effects are included for the quartic \betafunction  \cite{Degrassi:2012ry}, then the 5-4-3 counting should be adopted. This requires an additional theoretical effort to compute the gauge and Yukawa \betafunctions to the corresponding order.  The key point for any renormalization group analysis, as shown above, is that the \betafunctions are linked through $\tilde{a}$.  This implies that any perturbative truncation made at the level of $\tilde{a}$ will  be consistent. Conversely if the truncation is made at the level of the \betafunctions unphysical features may well appear \cite{Antipin:2012kc}.

\section{Vacuum stability analysis}
\label{sec:321}

The analysis of the vacuum stability requires the knowledge of the effective potential of the model at hand.  The standard model effective potential is known up to two loops. Its explicit form is given in the appendix of Ref.~\cite{Degrassi:2012ry}. For large field values  $\phi\gg v=246$ GeV, the potential is very well
approximated by its RG-improved tree-level expression,
\begin{equation}
V_{\eff}^{tree}=\frac{\lambda(\mu)}{4}\phi^4 \ .
\end{equation}
with $\mu=\mathcal O({\phi})$. Therefore if one is simply interested in the condition of absolute
stability of the potential, it is possible to study the RG evolution of $\lambda$ and determine the largest scale $\Lambda<M_{pl}$, with $M_{pl}$ the Planck scale, above which the coupling becomes negative.

We now compare the RG evolution of the standard model Higgs quartic coupling within the 3-2-1 Weyl consistent counting to the 3-3-3  counting.\footnote{ For the 3-3-3 counting scheme we use the
state-of-the-art three-loop standard model \betafunctions Refs.\cite{Holthausen:2011aa,Bednyakov:2012en, Chetyrkin:2012rz, Chetyrkin:2013wya, Bednyakov:2013eba, Mihaila:2012fm}.} The RG evolution of the standard model Higgs-self interaction coupling in both counting schemes is shown in Fig.~\ref{fig:1a}, where we used the PDG value for the top mass $M_t=173.5\pm 1.4$~GeV \cite{Beringer:1900zz} and the CMS measurement of the Higgs mass, $M_H = 125.7 \pm 0.6$~GeV \cite{CMS-PAS-HIG-13-005}. We observe that in both counting schemes $\lambda$ crosses zero at the scale $\Lambda\approx 10^{10}$ GeV, although the crossing happens at a slightly lower scale in the 3-2-1 counting. 

\begin{figure}[bt]
\subfloat[RG evolution of the standard model Higgs quartic coupling. We indicate with $\lambda_{333}$ ($\lambda_{321}$) the evolution of the $\lambda$ coupling according to the 3-3-3 (3-2-1) scheme. ] 
{\includegraphics[width=.48\textwidth]{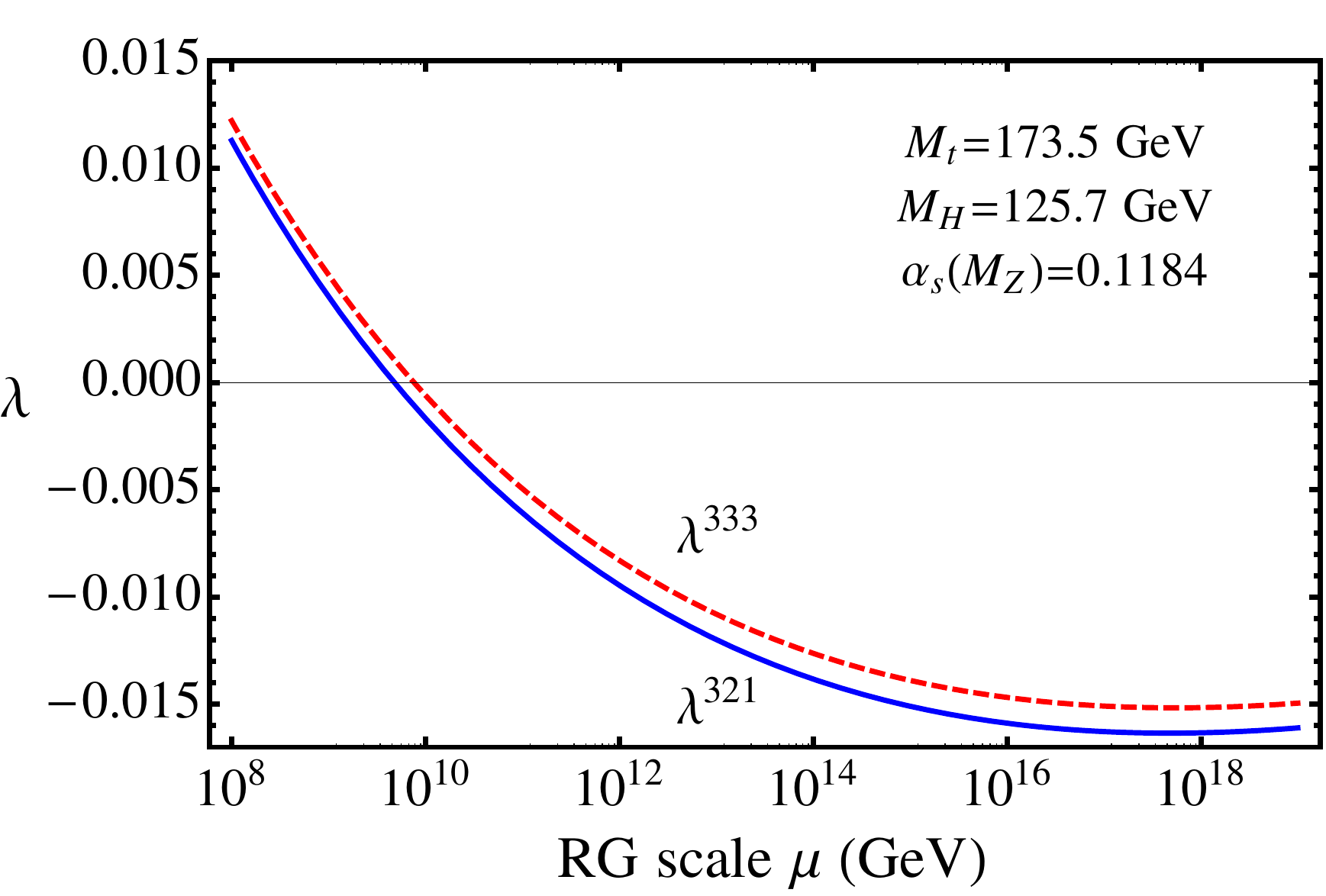}\label{fig:1a}}
  \hfill
\subfloat[RG evolution of the effective standard model Higgs quartic coupling. We indicate with $\lambda_{\rm eff}^{333}$ ($\lambda_{\rm eff}^{321}$) the evolution of the $\lambda_{\rm eff}$ coupling according to the 3-3-3 (3-2-1) scheme.]{\includegraphics[width=.49\textwidth]{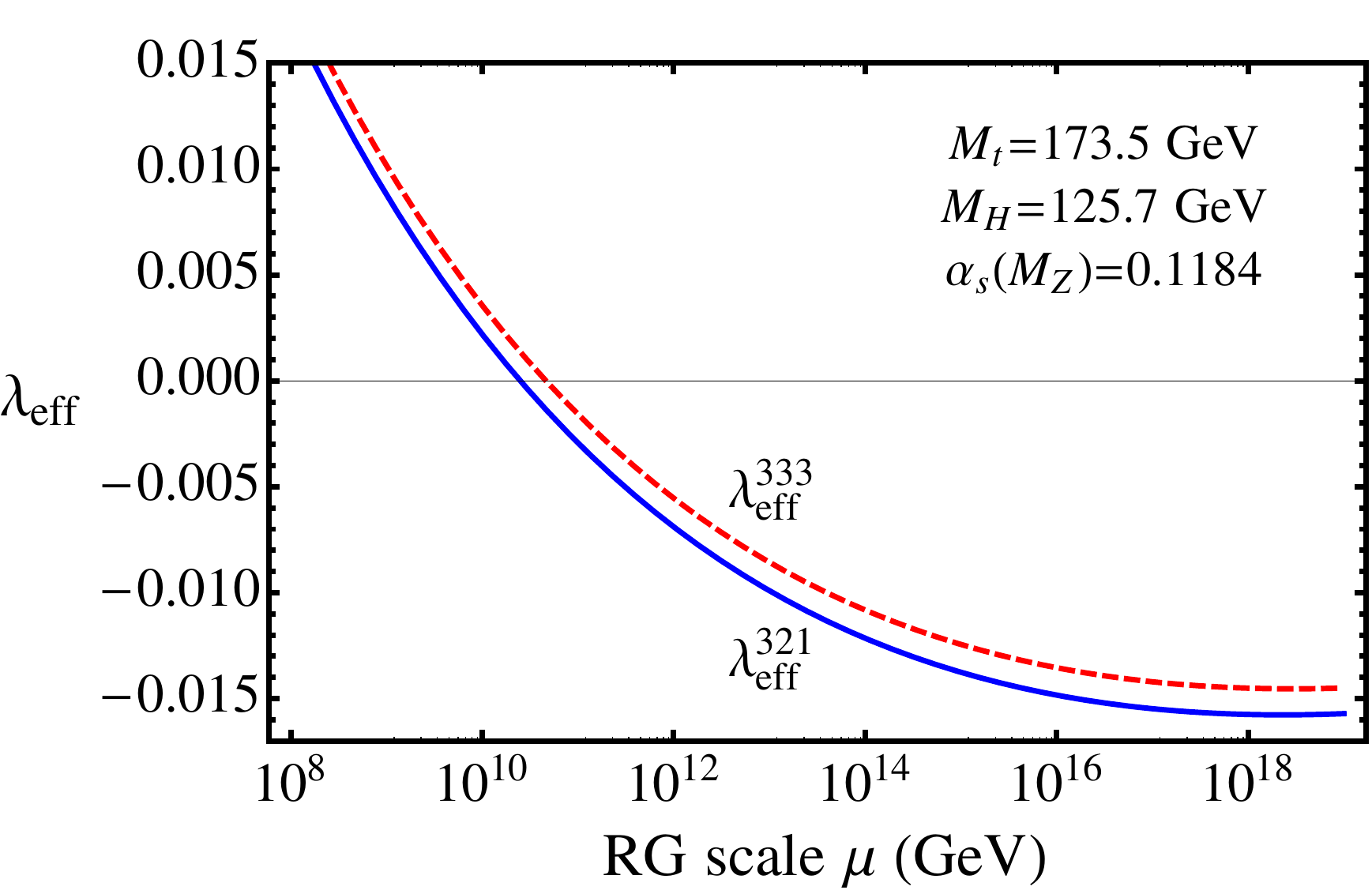}\label{fig:1b}}
\addtocounter{figure}{1}
\end{figure}

However, an accurate determination of the scale $\Lambda$ has to take into account the full structure of the Higgs potential. As was shown in \cite{Casas:1994qy,Casas:1994us}, one can always define an effective coupling $\lambda_{\eff}$ such that for $\phi\gg v$ the effective potential assumes the form
\begin{equation}
V_{\eff}=\frac{\lambda_{\eff}(\mu)}{4}\phi^4 \ .
\end{equation}
The explicit expression for $\lambda_{\eff}$, up to two-loop order, can be found in \cite{Degrassi:2012ry}. Within the 3-2-1 counting scheme, we have to take into account $\lambda_{\eff}$ only to one-loop order, which is consistent with the one-loop running of the quartic coupling. On the other hand for the 3-3-3 scheme we keep the full two-loop expression. The direct comparison between the running of the effective quartic couplings in the two schemes is shown in Fig.~\ref{fig:1b}. We note a pattern very similar to the one for $\lambda$ given in Fig.~\ref{fig:1a}. The  difference is, however, that the scale where $\lambda_{\eff}$ crosses zero is roughly one order of magnitude larger, $\Lambda\approx 10^{11}$GeV.

\begin{figure}[bth]
\includegraphics[width=.7\textwidth]{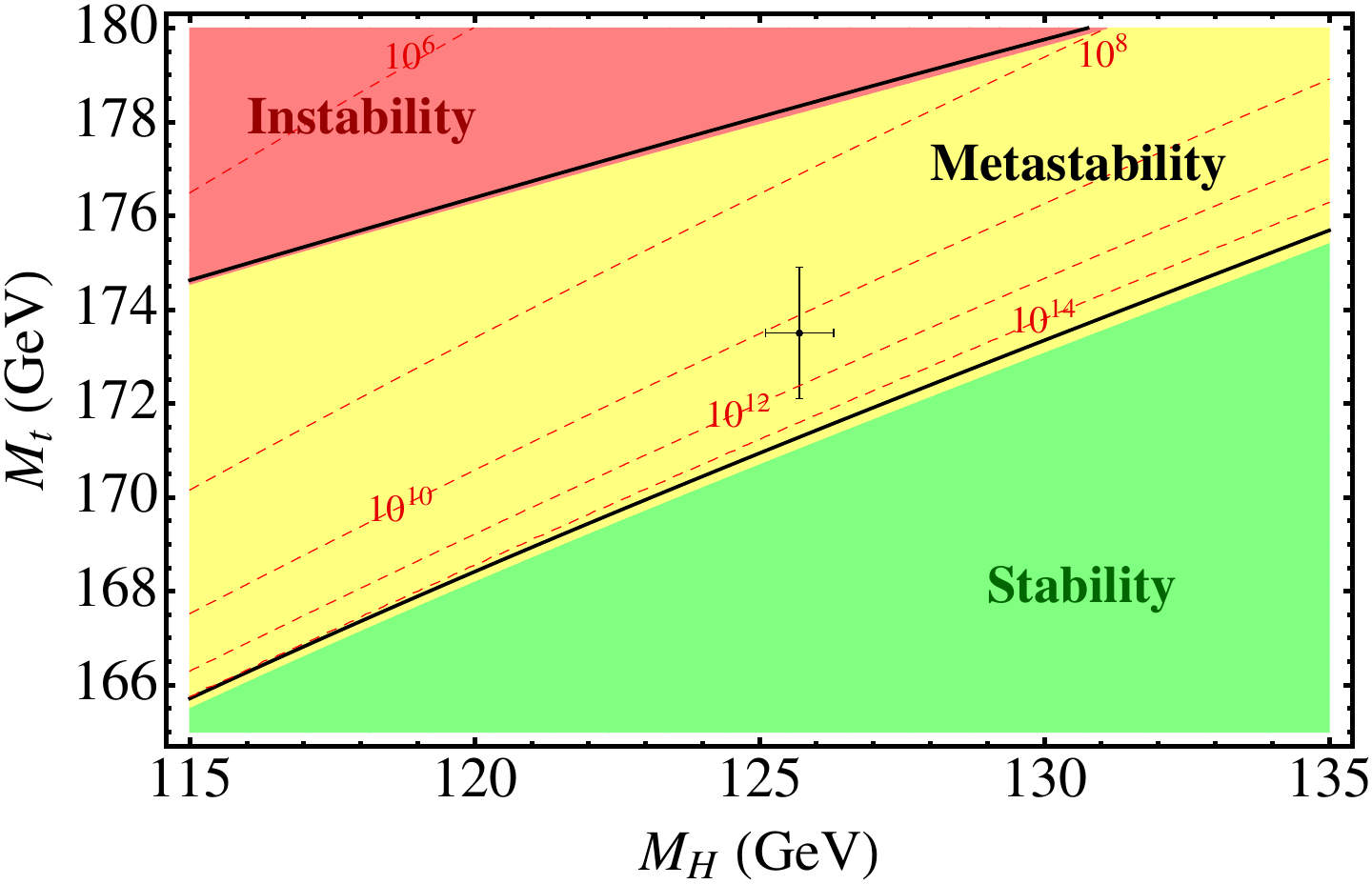}
\caption{Standard model stability analysis based on the effective standard model Higgs quartic coupling. The red region indicates instability, the yellow metastability and the green absolute stability following the 3-2-1 counting. For comparison, the black lines indicate the bounds from the 3-3-3 counting. The point with error bars shows the experimental values of the top \cite{Beringer:1900zz} and Higgs \cite{CMS-PAS-HIG-13-005} masses. The red dashed lines show the value in GeV at which $\lambda_\eff^{321}$ crosses zero.}\label{fig:moneyPlot}
\end{figure}

We have also studied the possibility that the standard model is in a metastable vacuum that may in principle decay at a later time. However, if the time it takes for the vacuum to decay is longer than the lifetime of the universe, this is not of immediate concern. To illustrate the situation we have plotted the stability of the standard model as a function of the top and Higgs masses (see Fig.~\ref{fig:moneyPlot}). The criterion for stability is that the quartic coupling is positive at least all the way to the Planck scale. On the other hand, for metastability we must require that the probability (with certain standard approximations, see \cite{Isidori:2001bm} for details) of the false vacuum decaying within the lifetime of the universe is less than one. This can be expressed mathematically as
\begin{align}
 \lambda(\phi)>-\frac{8\pi^2/3}{4 
 \log [\phi T_U e^{\gamma_E}/2]},
\end{align}
where $T_U$ is the age of the universe and $\gamma_E$ is the Euler-Mascheroni constant.

\begin{figure}[bt]
\subfloat 
{\includegraphics[width=.48\textwidth]{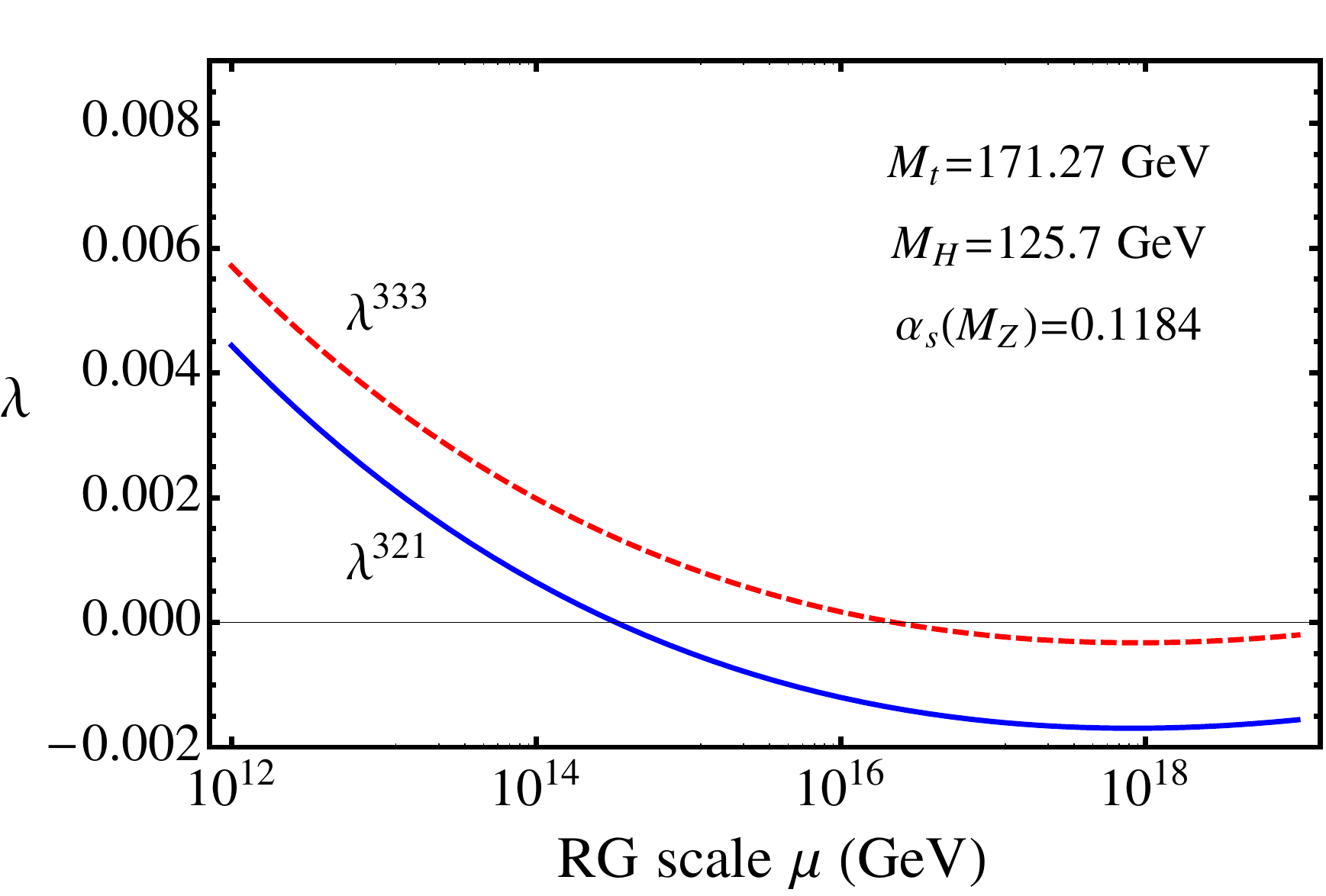}\label{fig:2a}}
  \hfill
\subfloat {\includegraphics[width=.49\textwidth]{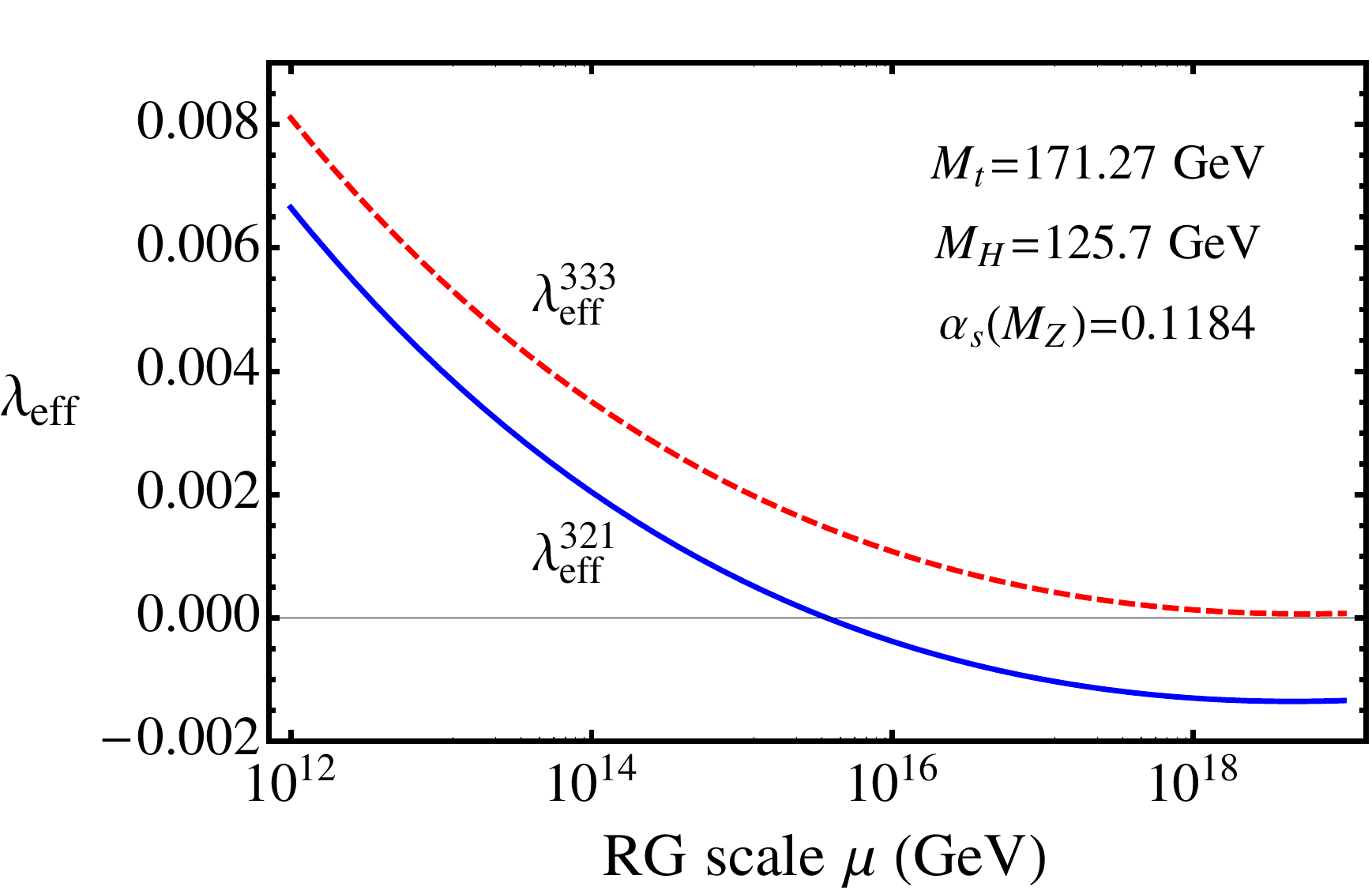}\label{fig:2b}}
\caption{RG evolution of the (effective) standard model Higgs quartic coupling. The mass of the top is tuned such that for $\lambda^{333}_{\rm eff}$ the potential develops a minimum at high energy, which is degenerate with the electroweak one. }
\label{fig:2}
\end{figure}

In addition to the vacuum stability analysis, we consider the case where the electroweak vacuum is the true ground state, but an unstable minimum exists  at higher values of the Higgs field. 
The condition for such a second vacuum close to the point when  $\lambda_{\eff}$ vanishes is the simultaneous  vanishing of $\beta_{\eff} = d\lambda_{\eff}/d\ln \phi  $ on the new minimum.  Typically these two conditions are met by lowering the value of the top mass.  To verify this possibility we show in the left and right panels of Figs.~\ref{fig:2} the evolution of the quartic couplings, as done in Figs.~\ref{fig:1a} and \ref{fig:1b}, but adopting a lower value of the top mass, i.e. $M_t=171.27$ GeV.  It is clear from the picture, that for this value of the top mass and within the  3-3-3 counting scheme, the conditions for the existence of a second vacuum, degenerate in energy with the electroweak one, are met. Indeed, in the right panel of Fig.~\ref{fig:2} we observe that $\lambda_{\eff}^{333}$ crosses zero at $\Lambda\approx 10^{19}$ GeV  with a near zero slope, i.e. $\beta_{\eff} \approx 0$. However, within the 3-2-1 counting scheme, the situation differs as $\lambda_{\eff}^{321}$ crosses zero about three orders of magnitude earlier, with non-vanishing $\beta_{\eff}$, for the same value of the top mass. We have to substantially lower the top mass to circa  $M_t\approx 171.05$ GeV  in this Weyl consistent scheme to accommodate the emergence of a degenerate minimum, giving a deviation of the order 2$\sigma$ from the central value of  the top mass.

\section{Conclusions}
\label{sec:conclusions}
We introduced a counting scheme for the organisation of the standard model perturbative expansion preserving the Weyl consistency conditions. These important conditions stem from conformal invariance which is a property of the standard model at energies higher than the electroweak scale. They non-trivially relate the various \betafunctions of the  theory. We  briefly reviewed the derivation and relevance of these conditions and defined the proper counting scheme. 

As a phenomenologically relevant example we investigated the vacuum stability of the standard model, by studying the running of its couplings up to the Planck scale within the new counting scheme. We showed that while the effects on the absolute stability of the model are small, sizeable effects appear when investigating the possible existence of a new unstable Higgs vacuum at high energies. 

Our results show that it is crucial for estimating theoretical uncertainties to consistently go to the next-to-leading order in all of the couplings, corresponding to a 4-3-2 counting. With the current state-of-the-art computations this only requires the derivation of the four-loop gauge \betafunctions.

\subsection*{Acknowledgements}
The authors would like to thank Diego Becciolini and Marco Nardecchia for stimulating discussions.
The CP$^3$-Origins centre is partially funded by the Danish National Research Foundation, grant number DNRF90.

\bibliographystyle{apsrev4-1}
\bibliography{biblio}

\end{document}